\documentclass[aps,prl,showpacs,nofootinbib,floatfix,twocolumn]{revtex4}
\usepackage{bm}
\usepackage{latexsym}
\usepackage{dcolumn}
\usepackage{amsfonts,amssymb,amsmath}
\usepackage{graphicx,epsfig}
\usepackage{psfrag}
\usepackage{subfigure}
\usepackage{feynmf}
\usepackage{rotating}
\usepackage{hyperref}
\usepackage{color}
\hypersetup{
    bookmarks=true,         
    unicode=false,          
    pdftoolbar=true,        
    pdfmenubar=true,        
    pdffitwindow=false,     
    pdfstartview={FitH},    
    pdftitle={My title},    
    pdfauthor={Author},     
    pdfsubject={Subject},   
    pdfcreator={Creator},   
    pdfproducer={Producer}, 
    pdfkeywords={keyword1} {key2} {key3}, 
    pdfnewwindow=true,      
    colorlinks=true,       
    linkcolor=red,          
    citecolor=cyan,        
    filecolor=magenta,      
    urlcolor=green,           
    linktocpage=true
}

\def\beq{\begin{equation}}
\def\eeq{\end{equation}}
\def\br{\begin{eqnarray}}
\def\er{\end{eqnarray}}
\def\benu{\begin{enumerate}}
\def\efnu{\end{enumerate}}
\def\nn{\nonumber}

\def\cl{{\cal C}_{\ell}}

\begin{document}
\title{Whipped Inflation : Inflation with suppressed scalar power spectra}
\author{Dhiraj Kumar Hazra}\email{dhiraj@apctp.org} 
\affiliation{Asia Pacific Center for Theoretical Physics, Pohang, Gyeongbuk 790-784, Korea}
\author{Arman Shafieloo}\email{arman@apctp.org}
\affiliation{Asia Pacific Center for Theoretical Physics, Pohang, Gyeongbuk 790-784, Korea\\ Department of Physics, POSTECH, Pohang, Gyeongbuk 790-784, Korea}
\author{George F. Smoot}\email{gfsmoot@lbl.gov}
\affiliation{Lawrence Berkeley National Laboratory, Berkeley, CA 94720, USA\\Institute for the Early Universe, Ewha Womans University Seoul, 120-750, Korea\\Paris Centre for Cosmological Physics, APC, Universite Paris Diderot, France}
\author{Alexei A. Starobinsky}\email{alstar@landau.ac.ru} 
\affiliation{Landau Institute for Theoretical Physics RAS, Moscow, 119334, Russia}

\begin{abstract}
 Motivated by the idea that inflation occurs at the GUT symmetry breaking scale, in this paper we construct a 
new class of large field inflaton potentials where the inflaton starts with a power law potential; 
after initial period of relative fast roll that lasts until after a few {\it e-folds} inside the horizon, it transits 
to the attractor of the slow roll part of the potential with a lower 
power. Due to the initial fast roll stages of inflation, we find a suppression in scalar primordial power at large scales and at 
the same time the choice of the potential can provide us a tensor primordial spectrum with high amplitude. This suppression in 
scalar power with a large tensor-to-scalar ratio helps us to reconcile the Planck and BICEP2 data in a single framework. 
We find that a transition from a cubic to quadratic form of inflaton potential 
generates an appropriate suppression in power of scalar primordial spectrum that provides significant improvement in fit compared to 
power law model when compared with Planck and BICEP2 data together. We calculate the extent of non-Gaussianity, specifically, the bispectrum 
for the best fit potential and show that it is consistent with Planck bispectrum constraints.
\end{abstract}
\maketitle
{\bf Introduction:~}Detection of B-mode polarization of Cosmic Microwave Background (CMB) photons by BICEP2~\cite{BICEP2:Detection,BICEP2:datasets} constrains 
the primordial tensor amplitude and thereby confirms yet another prediction of inflation. The constraints on tensor amplitude obtained through BICEP2 data brings in a new question 
in primordial cosmology since the bounds on the tensors obtained from Planck temperature anisotropy~\cite{Ade:2013uln,Planck:lilelihood} data have only 
limited overlap with the BICEP2 data. This inconsistency between the two datasets can be viewed from another interesting perspective were we can address these two data 
simultaneously. It has been showed in a number of recent papers~\cite{Hazra:recent,Contaldi:2014zua,Abazajian:2014tqa,Miranda:2014wga,Ashoorioon:2014nta} that a drop in large scale scalar primordial power spectrum (PPS) allows 
tensor-to-scalar ratio, $r$ to be large and helps to fit the Planck and BICEP2 data simultaneously and significantly better than power law scalar PPS. Using simple phenomenological models from~\cite{Hazra:2013nca},
in a recent paper~\cite{Hazra:recent}, we have shown that the addition of BICEP2 data rules out the power law form of the primordial spectrum at 
more than 3$\sigma$ CL. 

The above results motivate us to build inflationary models that can generate a suppression in scalar power at large scales along with 
reasonable amplitude of tensor PPS ($r\sim0.2$). Moreover, keeping the primordial bispectrum constraints from Planck in mind  
we should ensure that the potential do not generate large non-Gaussianity 
for a large window of cosmological scales. In this paper, staying within the canonical Lagrangian for single scalar field, keeping 
up to renormalizable terms in the inflaton potential, using the standard Bunch-Davies vacuum 
initial conditions for perturbations, we provide an inflationary potential which can address all the issues and fits the data
from Planck and BICEP2 significantly better than the power law scalar PPS. Whipped inflation signifies the scalar power spectra generated 
in this models resembles to a long snapped whip.

The paper is organized as follows. In next section we shall describe the form of an inflaton potential that we propose and use here. 
Afterwards we shall briefly discuss the numerical methods used to solve the potential and to compare the scalar and the tensor PPS 
to the data. In results and duscussions section we shall present the status of Whipped inflation in the light of Planck 
and BICEP2 data and finally we shall conclude with the main features of this analysis.

{\bf Inflationary potential:~\label{sec:formalism}}The potential we are proposing in this paper consists of a rapidly varying and a slowly varying part (Eq.~\ref{eq:potential}). Afterwards, in this 
section we provide the motivation of the construction of this potential.

\beq
V(\phi)=V_S(\phi)+ V_R(\phi),~\label{eq:potential}
\eeq

The basic construction our potential depends on two facts. Firstly, to generate large $r$ we need to work with large field models. 
Secondly since the largest scale modes leave the Hubble radius in the earliest times, an initial brief period of relative fast roll (fast compared to 
the next slow roll phase) can help suppressing the large scale power and at the same time producing higher tensor mode perturbations.

In this paper we choose to work with the following rapid and slow part of the potential as in Eq.~\ref{eq:potential_rs}. 

\begin{eqnarray}
V_S(\phi)&=&\gamma\phi^p\nn\\
V_R(\phi)&=&\lambda(\phi-\phi_0)^q~\Theta(\phi-\phi_0),~\label{eq:potential_rs} 
\end{eqnarray}
where, $\Theta(\phi-\phi_0)$ denotes the Heaviside theta function which cuts off the contribution of the rapid part beyond $\phi\le\phi_0$.
Note that both $V_S(\phi)$ and $V_R(\phi)$ contain power law potentials. We start near the minima of the potential $V_R(\phi)$, {\it i.e.}
near $\phi=\phi_0$ (but still for $\phi_{\rm initial}>\phi_0$ and $\phi_{\rm initial}-\phi_0 \sim 3-4~{\rm M_{PL}}$) where the field rolls relatively fast 
(yet $\epsilon_{\rm H}=-\dot{H}/H^2<1$) and then reaches the attractor of $V_S(\phi)$. Since the inflaton $\phi$ will 
reach $V_S(\phi)$ after a initial fast roll period, the power $p$ of the slowly varying part will determine the amplitude of the tensor
PPS at smaller scales. $V_R(\phi)$ on the other hand defines the suppression at large scales. 

Similar type of transition was originally used in~\cite{S92} for $q=1$, however since $r$ was a free parameter for this model, 
so it could not be predicted in advance~\footnote{A similar PPS was also studied in~\cite{CPKL03}.}. For $q=2$,
~\cite{JSS08,JSSS09} discuss the effects in primordial power spectrum in hybrid inflation 
scenarios where the potential generates a step in the spectral index, modulated by characteristic oscillations. 
It was recently generalized to an arbitrary $q$ in~\cite{Bousso:2013uia} basing on the hypothesis of the first order phase 
transition in the inflaton field at the GUT scale with formation of Coleman - de Luccia bubbles, following the previous papers 
on this topic~\cite{Linde:1998iw,Linde:1999wv}, and similar to what was originally proposed in~\cite{Sato:1980}, but
followed by $N\sim 60$ e-folds of more standard slow-roll inflation. 
However, this picture is not the only possibility. Alternatively, as was assumed in~\cite{S92,JSS08,JSSS09}, such 
potential may arise due to a fast phase transition in {\em another} massive field coupled to the inflaton. 
This transition is of the first order for $q=1$ and second order for $q=2$, see~\cite{S98}. 
In this case, slow roll inflation is only temporarily weakly broken around the moment of this transition.

In this paper, however, keeping in mind that we need to introduce a large scale suppression in scalar power for a wide range of cosmological
scales, we need initial and extended fast roll period with a smooth transition to the slow roll phase more than localized features. 
Hence the choice of $V_S(\phi)$ and $V_R(\phi)$ are most important here in order to generate appropriate suppression within the {\it single canonical
scalar field} model with the minimal number of extra parameters. In our analysis, we choose to work with $(p,q)=(2,3), (2,4)$ and $(3,4)$. 
Note that higher the $q$, the higher the tensor amplitude will be. We understand that assuming higher powers in $q$ in the second of Eq~\ref{eq:potential_rs}
requires more fine tuning, in other words, more amount of hidden symmetry of inflaton interactions with itself and other fields. 
However, our aim is to investigate phenomenological consequences of such an assumption. 
These values of $p$ and $q$ are chosen since they work reasonably fine and because they are suitable as one 
has some guides from attempts at Higgs induced inflation and vacuum stability on the Higgs potential.

At this point we should mention, since first year WMAP data, there were indications of large scale 
suppression in scalar power~\cite{Peiris:2003ff} and model independent reconstructions~\cite{reconstruction} too suggest the possibility 
of large scale suppression and different features in the scalar spectrum. Note that large scale suppression of scalar PPS due to an inflaton fast-roll stage prior to its slow-roll has been discussed in~\cite{TW04,K05,LB13}. Through intermediate fast roll during inflation, the suppression in power was addressed in 
different references~\cite{features}. In paper~\cite{Hazra:2010ve}, keeping the importance of tensor perturbations in mind, the complete scalar and tensor power spectra were confronted with different combinations of CMB datasets in canonical and non-canonical scalar field scenarios.

In this work too, a mild amount of temporal fast-roll is used in Whipped inflation. We should emphasize that according to the present demand of the Planck and BICEP2 data combination~\cite{Hazra:recent} and being within a renormalizable canonical scalar field theory, Whipped inflation is the first framework that can provide the overall suppression in scalar power, appropriate tensors and low non-Gaussianity.

{\bf Few comments on the numerical methods:~\label{sec:num}}We solve the background inflationary equations and the scalar and tensor 
perturbation equations using the publicly available code 
BI-spectra and Non-Gaussianity Operator, {\tt BINGO}~\cite{Hazra:2012yn}. We have fixed the initial conditions for inflaton by 
allowing sufficient {\it e-folds} $\sim70$ and by using initial slow roll. We have assumed that the pivot scale $k_{\ast}=0.05~{\rm Mpc^{-1}}$
leaves the Hubble radius 50 {\it e-folds} before the end of inflation. However, for $(p,q)=(3,4)$ case we have also worked with the assumption that 
$k_{\ast}$ leaves 60 {\it e-folds} before the inflation ends. We shall denote this number of {\it e-folds} by the usual convention $N_{\ast}$.
The tensor part is calculated using a modified version of the same code, yet to be publicly available. We have modified {\tt CAMB}~\cite{cambsite,Lewis:1999bs} to 
work with the {\tt BINGO} outputs directly. We have used the {\tt COSMOMC}~\cite{cosmomcsite,Lewis:2002ah} to find the best fit using Powell's 
BOBYQA method of iterative minimization~\cite{powell}. We have used the {\tt commander} and {\tt CAMspec} likelihood to estimate the low-$\ell$
and high-$\ell$ likelihood from Planck data~\cite{Planck:lilelihood}. We have used WMAP low-$\ell$ (2-23) E-mode polarization data~\cite{Hinshaw:2012fq}
(denoted as WP in results section). The complete BICEP2 likelihood is calculated using bandpowers for 9 bins for E and B mode polarization data. 
We should also mention here that to make our analysis robust, we have allowed 
the background cosmological parameters and the 14 Planck foreground nuisance parameters to vary along with the inflationary potential parameters.
To calculate the non-Gaussianity for this inflationary model, we again use {\tt BINGO} in the equilateral limit. 
Note that POLARBEAR B-mode polarization~\cite{Ade:2014afa} data can also help in order to constrain the cosmology better and for 
a complete parameter estimation we expect to include POLARBEAR data in a future analysis with Whipped inflation.
In all our analyses we have assumed spatially flat FLRW universe. We have defined ${\rm M_{PL}}^2=1/(8\pi G)$ and used $\hslash=c=1$ 
throughout the paper.

{\bf Results and discussions:~\label{sec:results}}In this section we provide the best fit results obtained for the potential in Eq.~\ref{eq:potential} for different choices of $p$ and $q$.
Table~\ref{tab:bestfits} contains the best fit values of the inflationary potential parameters and different cosmological parameters.
Best fit $\chi^2_{\rm eff}$ ($-2\ln{\cal L}$) are provided along with their breakdown in different datasets. The value of 
$-2\Delta\ln{\cal L}$ indicates the difference between the log likelihood obtained in a particular model and the power 
law scalar and tensor PPS (for power law best fit values, see~\cite{Hazra:recent}) when compared with Planck + WP + BICEP2 data combinations. 
Note that when we work with $p,q=2,3$ (cubic to quadratic transition), 
we get maximum improvement in likelihood (approximately 8). The higher 
tensor-to-scalar ratio ($r\sim0.15-0.25$) in all these models helps to fit BICEP2 data as good as (or better than) 
power law model. Importantly, the suppression at the large scale scalar PPS, originated from the initial fast roll phase, helps 
to fit the Planck data significantly better then the power law model. For quartic to quadratic transition ($p=2~{\rm and}~q=4$) 
we find that the improvement in fit decreases marginally from $p,q=2,3$. This decrease can be attributed to the shape of the 
transition, which indicates that along with the slow roll part of the potential, the initial {\it `type'} of fast roll part is also 
important in order to address the data better. When we allow transition to cubic potentials, {\it i.e.} $p=3$, we 
know that due to even higher tensor-to-scalar ratio ($r\sim0.2-0.25$ compared to quadratic potential, we are able 
to fit the BICEP2 data better than power law (where power law optimizes 
the likelihood between Planck and BICEP2) but the fit to Planck data becomes a bit worse~\footnote{Fit to Planck data gets worse, since : 
1. corresponding to the high $r$, the suppression is not enough which fits {\tt commander} worse, and, 2. this model produces 
more red tilt in the scalar PPS ($n_{\rm S}\sim0.95$) at small scales than demanded by Planck data}. Even though, this model provides an overall
4 improvement in fit compared to the power law model. The result for quartic to cubic transition with $N_{\ast}=60$ is also provided,
where we notice further improvement in likelihood compared to the same transition with $N_{\ast}=50$.

\begin{table}[!htb]
\begin{center}
\vspace{4pt}
\begin{tabular}{|c | c | c | c | c |}
\hline\hline
 \multicolumn{5}{|c|}{{\bf Inflation potential (Eq.~\ref{eq:potential}) and cosmological parameters}}\\
\hline
& $p=2,q=3$ & $p=2,q=4$ & $p=3,q=4$ & $p=3,q=4$\\
& $N_{\ast}=50$ & $N_{\ast}=50$  & $N_{\ast}=50$ & $N_{\ast}=60$\\

\hline

$\Omega_{\rm b}h^2$ & 0.02206&0.02208 & 0.02208&0.02206\\
\hline

$\Omega_{\rm CDM}h^2$ & 0.1189& 0.1193& 0.1198&0.1191\\
\hline

$100\theta$ &  1.041& 1.041  & 1.041&1.041\\
\hline

$\tau$ & 0.098 & 0.096& 0.097&0.085\\
\hline

$\gamma$ & $2.6\times10^{-11}$ & $2.6\times10^{-11}$&$1.5\times10^{-12}$&$9.4\times10^{-13}$\\
\hline

$\lambda$ & $1.1\times10^{-10}$ &$5.5\times10^{-11}$& $4.6\times10^{-11}$&$5.2\times10^{-11}$\\
\hline

$\phi_0$($\rm M_{\rm Pl}$) & 14.27 & 14.11 & 17.17&18.97\\

\hline
\hline\hline

$\Omega_{\rm m}$ & 0.31&0.31  & 0.315&0.31\\
\hline
$H_{0}$ & 67.6& 67.4& 67.25&67.5\\
\hline\hline
\multicolumn{5}{c}{$-2\ln{\cal L}$ [Best fit]}\\
\hline
{\tt commander} & -8.49 &-6.83  &-4.08&-5.33\\
 $\ell=2-49$ &  & & &\\
\hline
{\tt CAMspec} & 7796.45 &  7796.67& 7797.69&7797.76\\
$\ell=50-2500$ &  & & &\\

\hline
{\tt WP} & 2013.52 & 2013.46& 2013.1&2013.4\\
\hline
{\tt BICEP2} & 40.43 & 40.4& 38.9&39\\
\hline
Total & 9841.91 & 9843.7 &9845.61 &9844.83\\
\hline
$-2\Delta\ln{\cal L}$ & -7.67 &-5.88  & -3.97 &-4.57\\

\hline\hline
\end{tabular}
\end{center}
\caption{~\label{tab:bestfits} Best fit parameters for the Whipped inflaton potential~\ref{eq:potential} and the best fit cosmological parameters 
when compared with Planck + WP + BICEP2 data combination. The breakdown 
of best fit likelihood in different datasets are provided along with the difference in log likelihood compared to the best fit power law scalar PPS
model~\cite{Hazra:recent}. Table contains the best fit parameters for different choice of $p$ and $q$. Note that for $p=2,q=3$ we are able to provide 
a significant ($-2\Delta\ln{\cal L}\sim-8$) better fit to the data compared to power law scalar PPS (or equivalently scalar PPS from 
a strict slow roll inflation). For other choice of $p$ and $q$ we also get substantial improvement. Note that when we allow the $p=3$, 
due to higher tensor-to-scalar ratio, the model fits the BICEP2 data better, but it does not fit the Planck low-$\ell$ data equivalently compared to the 
other cases. However, for $N_{\ast}=60$ we find the model with $p=3,q=4$ is fitting the data marginally better than $N_{\ast}=50$.}
\end{table}

\begin{figure*}[!htb]
\begin{center} 
\resizebox{210pt}{160pt}{\includegraphics{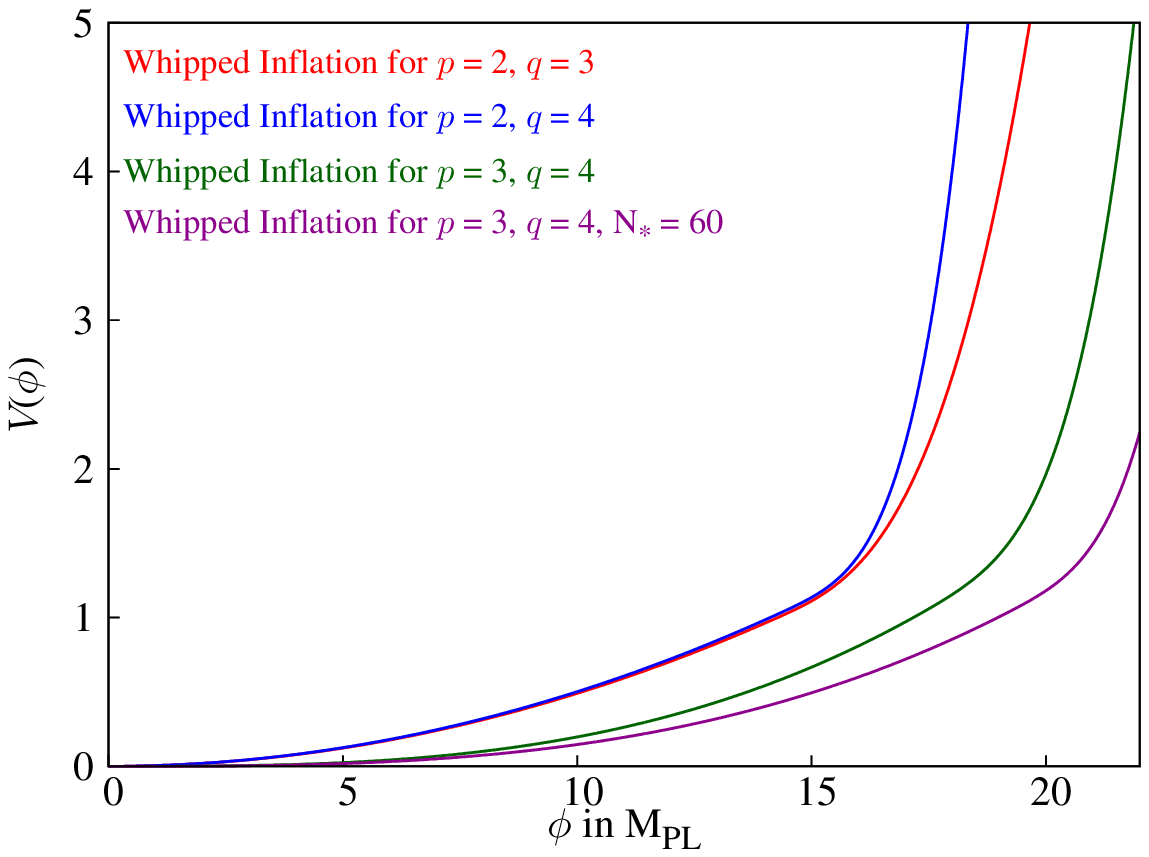}} 
\resizebox{210pt}{160pt}{\includegraphics{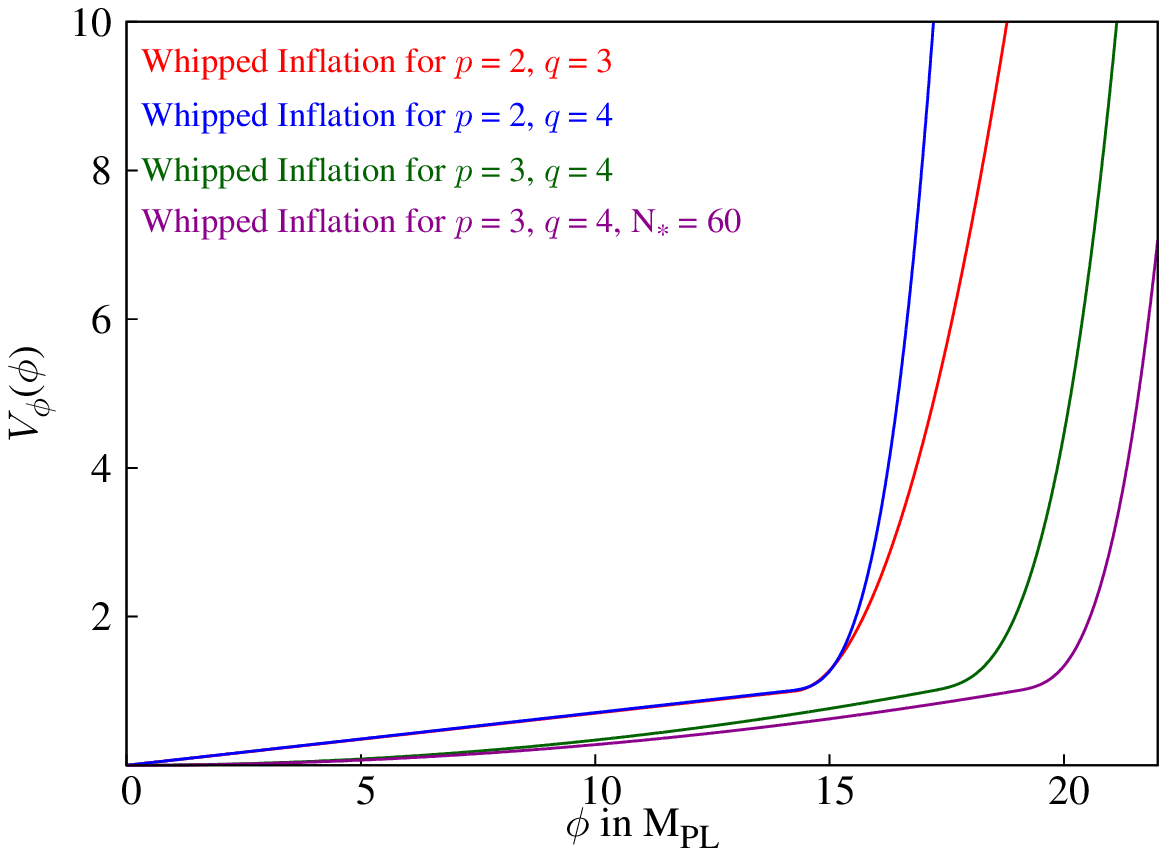}} 
\resizebox{210pt}{160pt}{\includegraphics{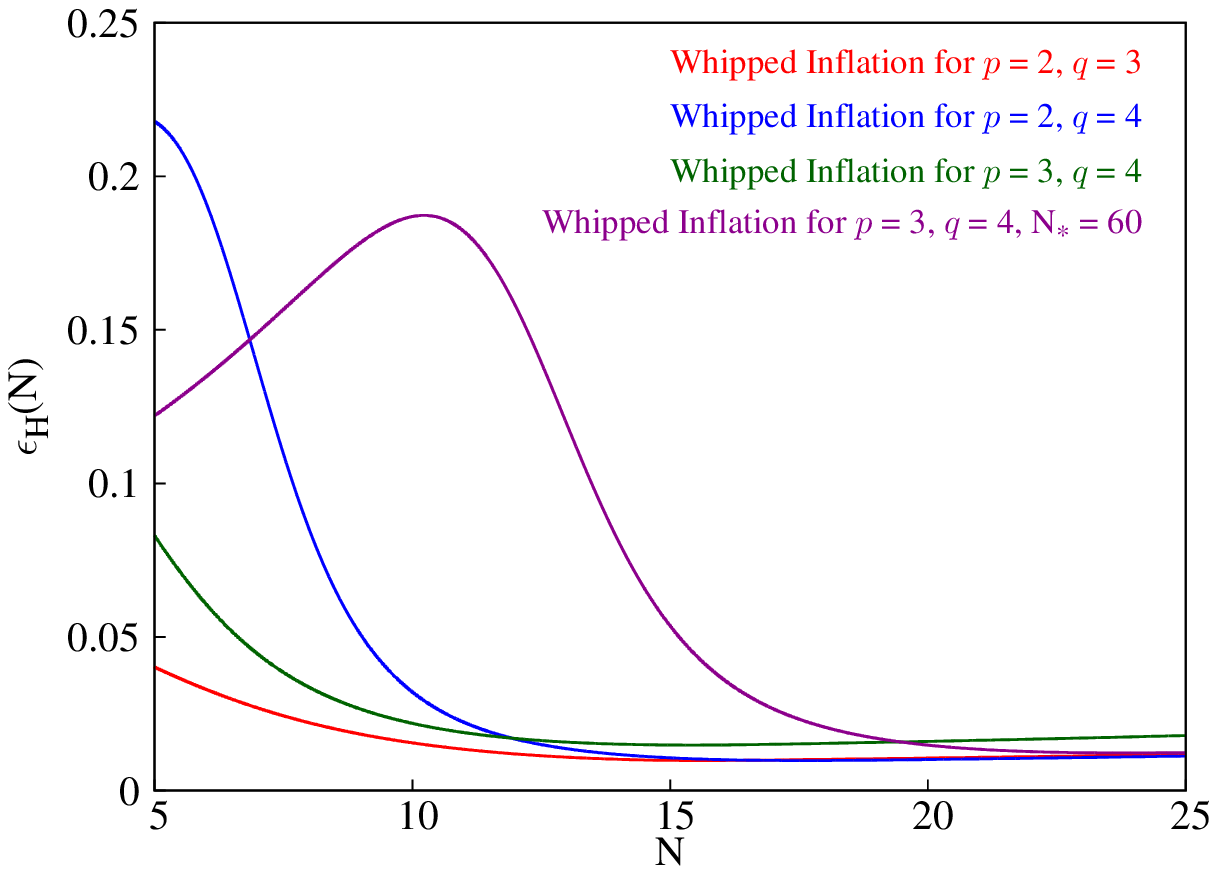}} 
\resizebox{210pt}{160pt}{\includegraphics{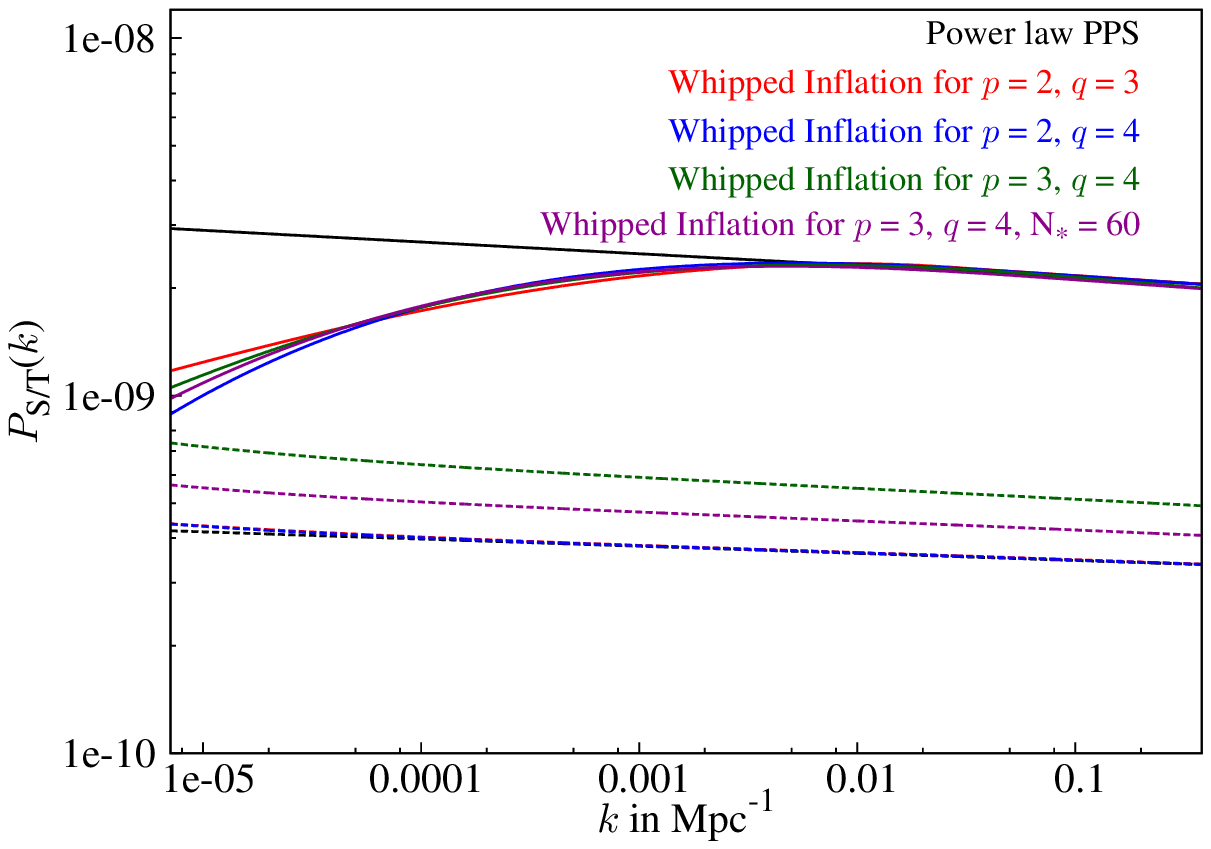}} 
\resizebox{210pt}{160pt}{\includegraphics{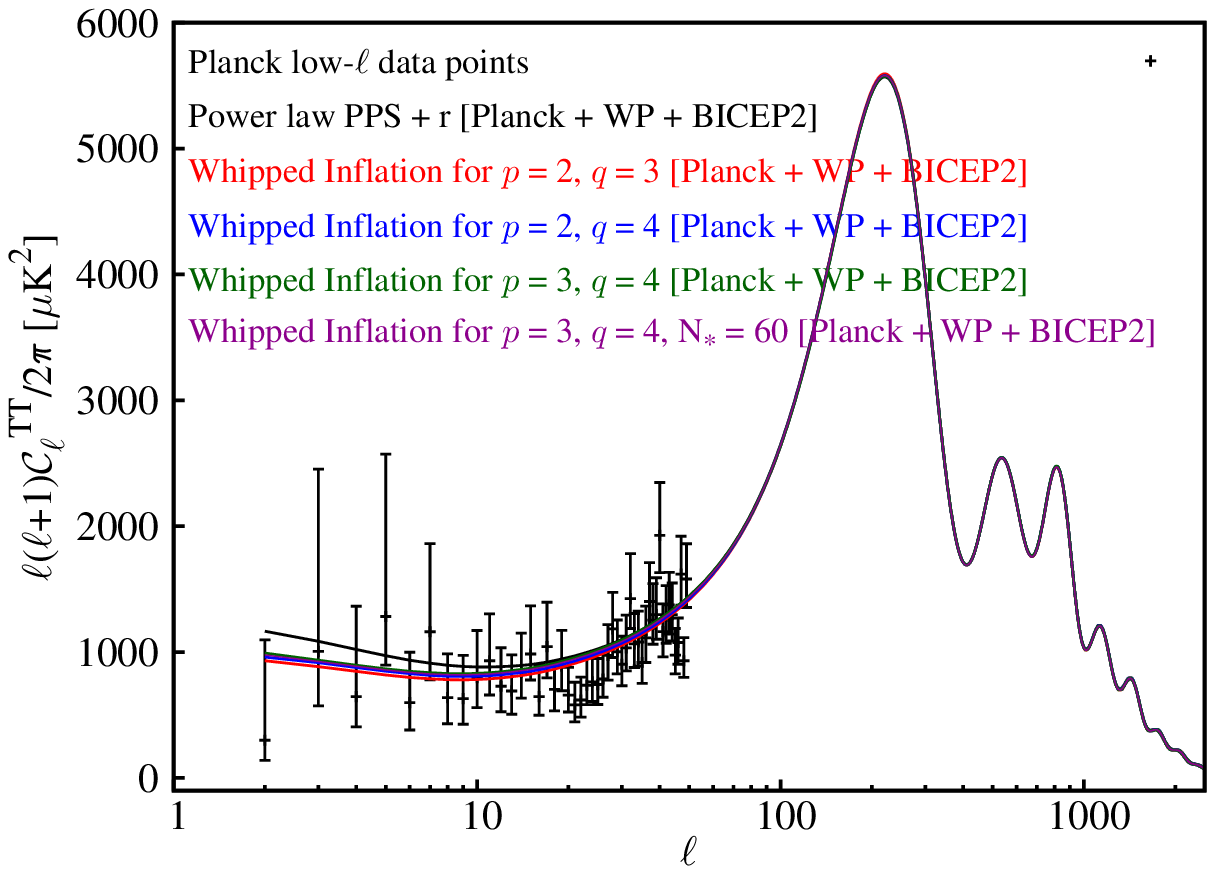}} 
\resizebox{210pt}{160pt}{\includegraphics{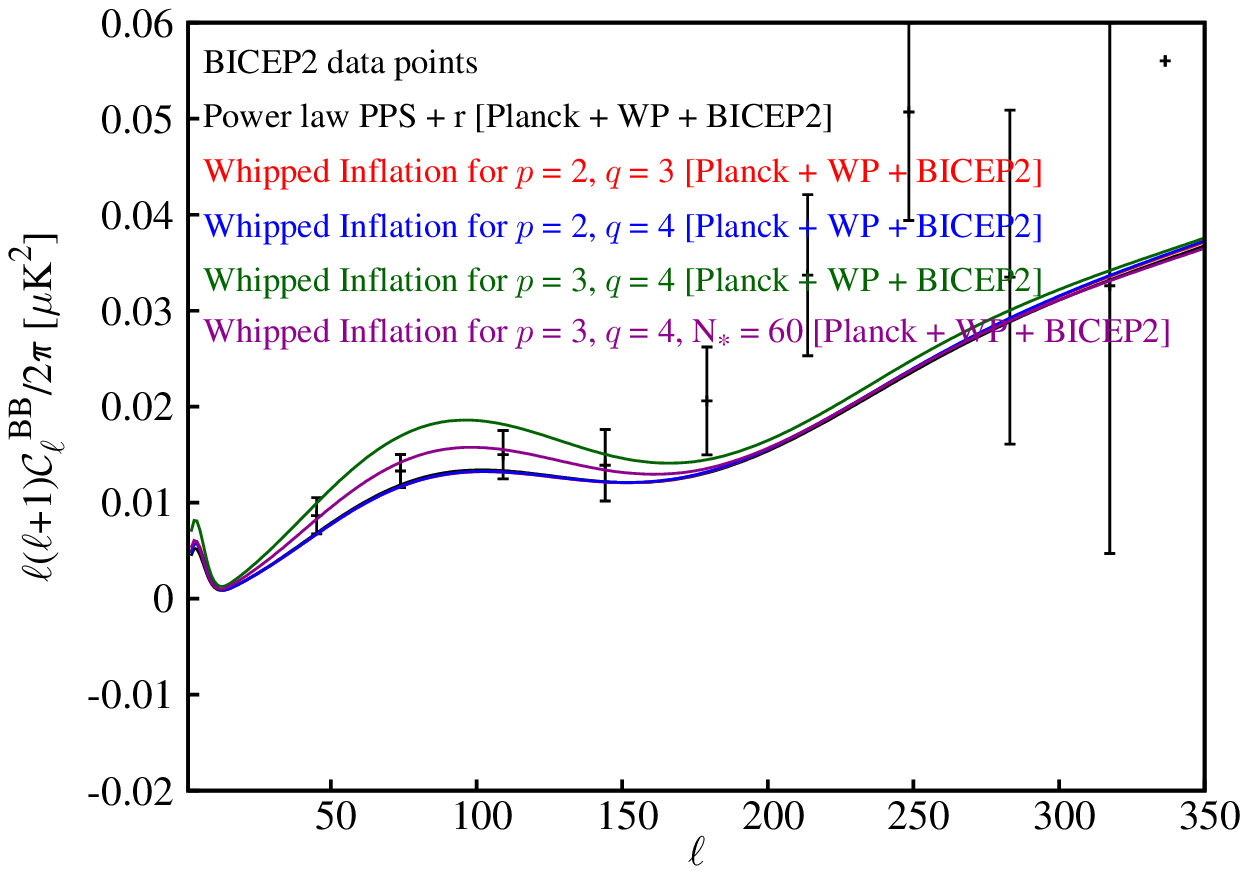}} 
\end{center}
\caption{\footnotesize\label{fig:theory}Top : Best fit inflaton potentials (left) and their derivatives (right) are plotted corresponding to
the best fit values in Table~\ref{tab:bestfits} obtained for Whipped inflationary potential (Eq.~\ref{eq:potential}). Middle left : First slow roll parameter, $\epsilon_{\rm H}=-\dot{H}/H^2$,
for different cases are plotted. Middle right : Best fit primordial scalar (solid) and tensor (dashed) power spectra are plotted along with 
power law best fit PPS (in black). Bottom : The best fit angular power spectra $\cl^{\rm TT}$ (left) and $\cl^{\rm BB}$ (right) are plotted
along with corresponding data points from Planck and BICEP2. The power law best fit spectra are plotted in black. Note that in all the cases,
our Whipped Inflation model fits Planck data significantly better than the power law through the suppression in large scale $\cl^{\rm TT}$.}
\end{figure*}

\begin{figure}[!htb]
\begin{center} 
\resizebox{210pt}{160pt}{\includegraphics{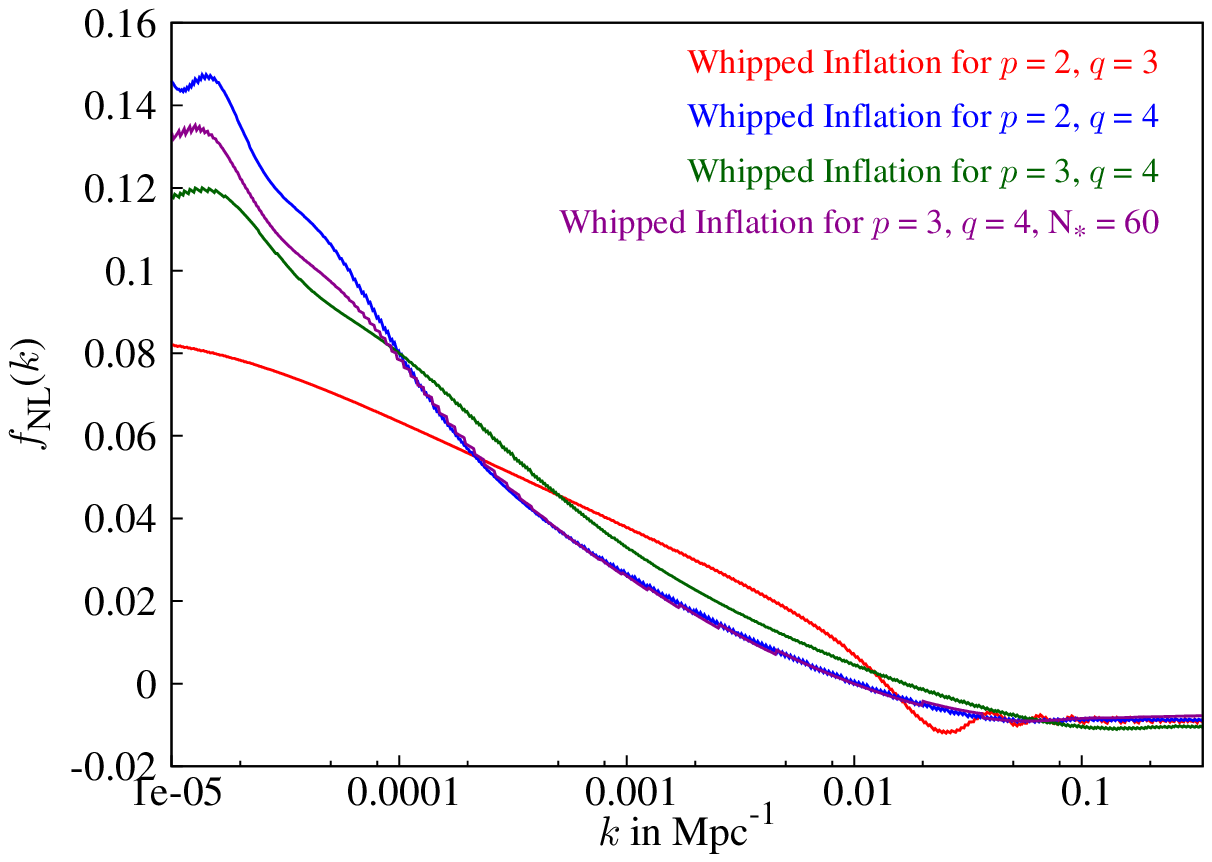}} 
\end{center}
\caption{\footnotesize\label{fig:fnltheory}The $f_{\rm NL}$ in equilateral triangular configurations ($k_1=k_2=k_3=k$) are plotted for best fit model are plotted. 
In all the cases, within cosmological scales, the bispectrum contribution is small and consistent with Planck limits.}
\end{figure}

In Fig.~\ref{fig:theory}, we plot the best fit results corresponding to the values of the parameters given in Table~\ref{tab:bestfits}.
We plot the best fit potentials and their derivatives and the first slow roll parameter $\epsilon_{\rm H}=-\dot{H}/H^2$. Note that the 
potentials and their derivatives are normalized to 1 at the point where inflaton transits to the slow roll phase, {\it i.e.} at $\phi=\phi_0$. 
The plot of slow-roll parameter $\epsilon_{\rm H}$ shows the fast roll to slow roll transition during initial inflationary epoch. For $p=3$ ($N_{\ast}=50$),
we find that the slow roll parameter settles down to a larger value as expected, indicating a higher $r$ than what we obtain from $p=2$. We also
plot the scalar ($P_{\rm S}(k)$) and tensor ($P_{\rm T}(k)$) PPS from different models and power law PPS. Note that the different kinds of 
suppression in scalar power for different models are evident from the plot. Finally in the same figure we plot the best fit $\cl^{\rm TT}$
and $\cl^{\rm BB}$ along with the corresponding data points from Planck (low-$\ell$) and BICEP2. Note that our models, for all given values of 
$p$ and $q$ are able to address the data. Note that the suppression in low-$\ell$ $\cl^{\rm TT}$ is the main factor that improves the likelihood
significantly compared to power law model, as has been tabulated in Table~\ref{tab:bestfits}.

In Fig.~\ref{fig:fnltheory}, we plot the non-Gaussianity, specifically the $f_{\rm NL}$, in equilateral triangular configurations, corresponding
to the best fit values quoted in Table~\ref{tab:bestfits}. The $f_{\rm NL}$ is calculated using the publicly available code {\tt BINGO}
using the methods described in~\cite{Hazra:2012yn,maldacena-2003,chen,Martin:2011sn}.
Due to initial fast roll phase we find {\it relatively} high $f_{\rm NL}$ (${\cal O}(0.2)$)
at large scales and at small scales the $f_{\rm NL}$ settles to a value closer to {\it zero}. 
In all the cases, the generated $f_{\rm NL}$ in these models are completely consistent with Planck constraints~\cite{Planck:fnl}.

{\bf Conclusions:~\label{sec:conclusions}}In this paper we provide an inflaton potential for canonical scalar field model 
that can address the CMB temperature 
and polarization data from Planck and BICEP2 significantly better than the power law model.  
The potential offers a rapid and a slow roll part 
and we show that the inflaton, after having fast roll for around 10-15 {\it e-folds}, eventually falls on the attractor of the slow roll 
part of the potential. The initial fast roll period introduces a large scale suppression in scalar PPS. 
Since we have shown in our recent paper~\cite{Hazra:recent} that using Planck temperature anisotropy data and BICEP2 
polarization data (mainly B-mode data), a large scale scalar power suppression 
rules out power law scalar PPS at more than $3\sigma$ CL, the models described in this paper serves as a suitable 
representative models for consistently addressing Planck and BICEP2 data together. 

The results in this paper indicates the following facts:
\begin{itemize}
 \item Staying within canonical scalar field model, working with only renormalizable terms and using the simple Bunch-Davies
 vacuum initial condition, it is possible to have a set of inflationary potential that can resolve the inconsistencies between
 Planck and BICEP2 data.
 
 \item The significant improvement in fit ($-2\Delta\ln{\cal L}\sim-8$) compared to the power law model (or equivalently a 
 strict slow roll model) certainly keeps these class of potential in a higher ground.
 
 \item In our recent paper~\cite{Hazra:recent}, we have argued that we need more flexibilities than a power law power spectrum 
 in order to explain CMB data better. The same statement can be translated in inflationary theory that we need more flexibilities 
 than what we have in a strict slow roll inflation. Our analysis justifies the addition of three extra parameters, namely, the 
 form of the rapid potential (described by $q$ and $\lambda$) and the scale of the transition from fast to slow roll {\it i.e.} $\phi_0$.
 
 \item The type of power suppression at large scale scalar PPS is important. Hence a proper 
 transition from potential power $q$ to $p$ is necessary.
 
 \item We do not need to construct theories to get a blue tensor PPS tilt to address the data.
 
 \item Since the feature (suppression) is not localized in a small window of cosmological scales, the suppressions that the potentials 
 offer do not fit statistical uncertainties in the data.
 
 \item The non-Gaussianity in these models are small and hence certainly favored by Planck.
\end{itemize}

We would like to close the paper by commenting that, if the value of tensor-to-scalar ratio $r$, obtained from B-mode polarization data 
from BICEP2 persists, the potentials discussed in this paper offer simple and at the same time important mechanisms to generate 
scalar and tensor PPS that can consistently address temperature and the polarization anisotropy data in single framework.
The Whipped Inflation models will then be worthy of an effort to relate them to the GUT symmetry breaking phase transition.

Just before we submit this paper to the journal there have been a paper~\cite{Liu:2014mpa} suggesting that the detected B-Mode polarization data by BICEP2 might be contaminated by the radiation from galactic radio loops and this could largely affect any cosmological conclusion. We should clarify that our results are based on the BICEP2 and Planck data assuming that they have had a good control of the systematics, taking in to account the effects of all foregrounds.


{\bf Acknowledgments:~}D.K.H. and A.S. wish to acknowledge support from the Korea Ministry of Education, Science and Technology, Gyeongsangbuk-Do and Pohang 
City for Independent Junior Research Groups at the Asia Pacific Center for Theoretical Physics. We also acknowledge the use of 
publicly available {\tt CAMB} and {\tt COSMOMC} in our analysis.
The authors would like to thank Antony Lewis for providing us the new {\tt COSMOMC} package that takes into account the recent 
BICEP2 data. We acknowledge the use of WMAP-9 data and from Legacy Archive for Microwave Background Data Analysis (LAMBDA)~\cite{lambdasite}, 
Planck data and likelihood from Planck Legacy Archive (PLA)~\cite{PLA} and BICEP2 data from~\cite{biceprepo}. A.S. would like to acknowledge
the support of the National Research Foundation of Korea (NRF-2013R1A1A2013795). A.A.S. thanks Prof. E. Guendelman for hospitality in 
the Ben-Gurion University, Beer-Sheva, Israel, during the period when this paper has been finished. A.A.S. was also partially supported 
by the grant RFBR 14-02-00894.


\end{document}